\documentclass[aps,amsmath,amssymb,prd,nofootinbib,reprint,longbibliography]{revtex4-1}
\pdfoutput=1 

\usepackage{graphicx}
\usepackage{hyperref}

\newcommand{\comment}[1]{}
\newcommand{\beq}[1]{\begin{equation}\label{#1}} 
\newcommand{\eeq}{\end{equation}}
\newcommand{\beqa}{\begin{eqnarray}}
\newcommand{\eeqa}{\end{eqnarray}}
\newcommand{\M}{\ensuremath{\mathcal{M}}}
\newcommand{\N}{\ensuremath{\mathcal{N}}}
\newcommand{\del}{\partial}
\newcommand{\Del}{\nabla}
\newcommand{\w}{\wedge}
\renewcommand{\b}{\bar}
\renewcommand{\t}{\tilde}

\begin{document}
\title{Variations on the Dirac string}
\author{Brad Cownden}\email{cowndenb@myumanitoba.ca}
\affiliation{Department of Physics \& Astronomy, University of Manitoba,
Winnipeg, Manitoba R3T 2N2, Canada}
\author{Andrew R. Frey}\email{a.frey@uwinnipeg.ca}
\affiliation{Department of Physics, University of Winnipeg, 515 Portage Ave,
Winnipeg, Manitoba R3B 2E9, Canada}
\affiliation{Department of Physics \& Astronomy, University of Manitoba
Winnipeg, Manitoba R3T 2N2, Canada}

\date{\today}

\begin{abstract}
Dirac's original solution of the nontrivial Bianchi identity for magnetic 
monopoles \cite{Dirac:1948um}, which redefines the fieldstrength along the 
Dirac string, diagonalizes the gauge and monopole degrees of freedom.
We provide a variant of the Dirac string, which we motivate
through a formal expansion of the Bianchi identity. We show how to use our
variant prescription to study monopole electrodynamics without reference to
a dual potential and provide some applications.
\end{abstract}

\maketitle

\section{Introduction}

Magnetic monopoles, while experimentally elusive \cite{1712.09849}, 
hold a special place in 
particle physics; they complete electric-magnetic duality, provide a reason
for charge quantization, and arise in the spontaneous breaking of many
grand unification models.  At the same time, there is a long tradition of
reformulating the description of the monopole's interaction with the
electromagnetic field
since Dirac's original work \cite{Dirac:1931kp,Dirac:1948um}:
\begin{itemize}
\item Monopoles arise as solitons of broken gauge theories; the microscopic
description as a semiclassical field configuration describes the monopole's
interactions but requires tracking the full degrees of freedom of that gauge 
theory.  While of course necessary for processes like monopole creation and
annihilation (see eg \cite{1607.07460}), it may be computationally 
excessive when an effective description is valid.  

\item In the absence of electric charges, the fieldstrength can be defined in 
terms of a dual potential $\star F\equiv d\t A$, which couples to monopoles
in the same way the vector potential couples to charges.  Of course, if
the fields of both charges and monopoles are of interest, the fieldstrengths
from both the potential and dual potential must be superposed in a 
``democratic'' formalism (which makes it ill-suited to the quantum mechanics
of charges and monopoles).  This type of approach has been advocated at least
since \cite{Rohrlich:1966zz}.

\item As pointed out originally in \cite{Wu:1976ge,Wu:1976qk}, the 
vector potential is not a globally defined function but a section of a fiber
bundle. When the fieldstrength has a nontrivial Bianchi identity, the
potential is defined in at least two coordinate patches; 
the potentials in the two
patches are related by a gauge transformation in the overlap of the two 
patches. While mathematically rigorous, treating the potential in this manner
mixes the gauge and monopole degrees of freedom because the overlap region
moves with the monopole.  This formalism also obscures the coupling
between the potential and monopole (for example, \cite{Wu:1976qk} switch to 
a dual potential formalism to derive the monopole's classical equation of
motion, while \cite{Hirata:1978gk} following \cite{Brandt:1976hk} 
combined Dirac's formalism below with
the rigorous gauge patching procedure). 

\item The potential in a single coordinate patch extends to all of spacetime
except for a half-line singularity extending from the monopole (Dirac's 
famous string singularity).  In \cite{Dirac:1948um}, Dirac noted that the
electromagnetic fieldstrength can be written in terms of a globally defined
potential and an extra term supported on the string singularity.  This 
approach separates the electromagnetic and monopole degrees of freedom and
elucidates the monopole-fieldstrength coupling. However, it suffers some
conceptual difficulties, including a singular fieldstrength along the 
(arbitrarily chosen) string and a constraint that electric charges cannot
intersect the Dirac string worldsheet.  Pragmatically, the semi-infinite
string can be awkward.

\end{itemize}

In this paper, we present a new hybridization of Dirac's string formalism with
the rigorous gauge patching prescription by allowing the Dirac string to 
end on an unphysical reference monopole, and we review the 
derivation of Dirac's string prescription from the dual potential
formalism with an emphasis on the origin of the string.  
We also advocate for a particular
configuration for the string and give a simple physical picture for its 
origin.  The fixed prescription for the string means that the string worldsheet
embedding coordinated depend on the monopole position along the entire
worldsheet, unlike Dirac's original formalism; 
our approach presents computational advantages in several 
circumstances.  The extension of our results to higher-dimensional 
monopole-like branes appears in the companion paper \cite{andrew} by one of
us.  Throughout this paper, we work in 4-dimensional Minkowski spacetime
for simplicity. We list our conventions, particularly signs, in an appendix.

\section{Dirac's string variables}

The Maxwell equations and Bianchi identities with magnetic sources are 
\beq{maxwell} d\star F =-\star j\ ,\  \ dF = -\star\t\jmath\ ,\eeq
where $j^\mu$ is the electric current and $\t\jmath^\mu$ the magnetic current.
For a single monopole, the current is
\beq{current}
\t\jmath^\mu =g \int_\M d\tau \del_\tau X^\mu(\tau)\delta^4(x-X) =
g u^\mu(t) \delta^3(\vec x-\vec X)\ , 
\eeq
where $g$ is the monopole charge, \M\ is the monopole's worldline, 
$\tau$ the worldline time, and $X^\mu$ the monopole position.  The second
equality follows in a fixed reference frame with a worldline static gauge
$\tau=t$; $u^\mu$ is the monopole's 4-velocity.

Dirac's key observation stems from the fact that any conserved current
(in Minkowski spacetime) can be written as the divergence of a two-form,
or $\star\t\jmath= d\star H$ for some $H$.  In fact, the field of an 
isolated monopole provides such an $H$ subject to the additional constraint
of the sourceless Maxwell equations.  Instead, following Dirac
\cite{Dirac:1948um}, we define
\beq{stringcoupling} G^{\mu\nu}(x) \equiv \int_\N \b dY^\mu\w\b dY^\nu
\delta^4(x-Y)\ ,\eeq
where \N\ is the worldsheet of a string with boundary $\del\N=\M-\M_*$
($\M_*$ is a so-far arbitrary worldline), $Y^\mu(\tau,\sigma)$ is the string
position at worldsheet coordinates $\tau,\sigma$ with $\sigma$
increasing from $\M_*$ to \M, and 
$\b d=d\tau\del_\tau+d\sigma\del_\sigma$ is the worldsheet exterior derivative.
This has divergence
\beqa (\star \, d \star G)^\mu &=& \int_\N \b dY^\nu\w\b dY^\mu
\del_\nu\delta^4(x-Y)\nonumber\\ 
&=& -\int_\N \b d\left[ \delta^4(x-Y)\b dY^\mu\right]\nonumber\\
&=& (1/g)\t\jmath^\mu -(1/g)\t\jmath_*^\mu\label{Gdiv}
\eeqa
in terms of the currents of monopoles along \M\ and $\M_*$.\footnote{An
additional sign enters at the last equality from the worldsheet orientation
as described in the appendix.\label{sign}}
As a result, the field strength defined as $F=dA-g\star G$ automatically 
solves the Bianchi identity for the dynamical monopole along $\M$ as long as
$A$ includes the potential for a monopole with reference worldline $\M_*$
(with appropriate gauge patching). 

It is worth pausing now to discuss the configuration of the string.  In
Dirac's formalism, the string is a dynamical object with no kinetic term
(in modern parlance, a tensionless string), so the worldsheet \N\ is 
completely arbitrary except for the specification of its 
boundary.\footnote{This arbitrariness reflects how a gauge transformation 
modifies where the potential of a monopole becomes singular (or more 
precisely the patching required for the potential), so the string can never
develop a tension, even quantum mechanically, for an unbroken gauge symmetry.}
Dirac furthermore locates the worldline $\M_*$ at spatial infinity, so the
string becomes semi-infinite.  However, we are free to specify the 
configuration of the string; a simple choice in a given reference frame
is to choose the reference monopole to be stationary ($\M_*$ at constant
position $\vec X_*$) and the string to be the straight line from $\vec X_*$
to the monopole position $\vec X(t)$ at each time.  We can build an 
arbitrary worldsheet in this way by then extending a Dirac string from 
$\M_*$ to another reference worldline $\M_{**}$, and so on, and then letting
the reference worldlines approach each other.  On the other hand, with 
a single line segment, the worldsheet embedding coordinates $Y^\mu$ depend on 
the monopole position at all worldsheet at all worldsheet points $\tau,\sigma$
unlike in Dirac's formalism.  We motivate this 
prescription for the Dirac string configuration in section \ref{s:delta}.

It is also instructive to consider the Dirac quantization of charge 
for different formalisms.  With distinct gauge patches, quantization enforces
the requirement that a single-value wavefunction in one coordinate patch 
remains single-valued after the gauge transformation in passing to another
patch.  In Dirac's original formalism, there is only a single gauge, but the
wavefunction picks up a phase under motion of the string equal to $g$ times
the electric flux through the surface swept out by the worldsheet --- 
the surface is noncontractible since the charge cannot intersect the 
worldsheet, as we see below.  Since the
surface swept out by the string is closed (including the point at infinity),
charge quantization is given by the condition that the wavefunction remain
single-valued if we sweep the string around the charge.  When the reference
monopole is at a finite position, both mechanisms for charge quantization
are possible.  In particular, if the string worldsheet is arbitrary but 
$\M_*$ is fixed, we can sweep the string in a closed surface around a charge,
as per Dirac.  With a set prescription for the worldsheet like discussed 
above, however, the worldsheet only moves if we move the reference position
$\vec X_*$, so the string sweeps out open surfaces, removing the quantization
condition.  On the other hand, the potential must be defined in patches around
the reference point, so single-valuedness of the wavefunction still leads
to charge quantization.

The Dirac string formalism also provides a direct means of finding the 
monopole equation of motion, which is otherwise carried out indirectly through
the dual potential formalism.  Including the string coupling $G$ in the
fieldstrength, the variation of the Maxwell action with respect to
the monopole and worldsheet positions is 
\beqa
\delta S_{Max} &=& \frac g4 \epsilon_{\mu\nu\lambda\rho} \int d^4x F^{\mu\nu}(x)
\delta G^{\lambda\rho}(x)\label{eom1}\\
&=& \frac g4 \epsilon_{\mu\nu\lambda\rho} \int d^4x \int_\N F^{\mu\nu}(x)
\left\{\b dY^\lambda\w \b dY^\rho \delta Y^\alpha
\del_{Y^\alpha}\delta^4(x-Y)\right.\nonumber\\
&&\left.+\left[\b d\delta Y^\lambda\w \b dY^\rho
+\b dY^\lambda\w \b d\delta Y^\rho\right]\delta^4(x-Y)\right\}\ .\nonumber
\eeqa
We can now restrict to a worldsheet integral, first converting the derivative
on the delta function to one with respect to $x^\alpha$ and integrating by 
parts. Then, with all partial derivatives now with respect to $Y$,
\beqa \delta S_{Max}&=&
\frac g2\epsilon_{\mu\nu\lambda\rho}\int_\N \left\{ \b d\left[F^{\mu\nu}(Y)
\delta Y^\lambda \b d Y^\rho\right]\vphantom{\frac 12}\right.\label{eom2}\\
&&\left.+\frac 12\del_\alpha F^{\mu\nu}(Y)\left(\b dY^\alpha\w
\b dY^\lambda \delta Y^\rho\right.\right.\nonumber\\
&&\left.\left. +\b dY^\lambda\w\b dY^\rho \delta Y^\alpha+
\b dY\rho\w\b d Y^\alpha\delta Y^\lambda\right)\vphantom{\frac 12}
\right\}\ .\nonumber
\eeqa
The latter lines of (\ref{eom2}) reorganize to 
\beq{eom3}
\frac g2\int_\N (d\star F)_{\nu\lambda\rho}\b dY^\nu\w \b dY^\lambda \delta Y^\rho
\ ,\eeq
which would yield an interaction between the (unphysical) string --- 
and therefore the monopole --- 
with the electric current when the gauge fields are on shell. 
To avoid this unphysical result, Dirac imposed the additional condition that
charges not intersect the string.  The first term, on the other hand, 
gives an integral over \M\ (assuming $\M_*$ fixed),\footnote{with sign
determined per footnote \ref{sign}.} so it combines with 
the variation of the monopole's kinetic term to give the magnetic Lorentz 
force equation
\beq{lorentz} \del_\tau p_\mu =-g(\star F)_{\mu\nu}\del_\tau X^\nu\ .\eeq
Meanwhile, the electric charge couples to the redefined potential $A$, so 
its equation of motion turns out as usual except for contact terms with the
Dirac string, which are forbidden due to Dirac's condition.

It is important to note that our derivation of the monopole equation of motion
treated the potential $A$ as an independent degree of freedom in contrast
to the case without a Dirac string and with gauge patches. In that case, 
there is no coupling between the monopole and gauge field, only a hidden
dependence of $A$ on $X^\mu$ which should be treated as an explicit dependence.
The Dirac string removes this explicit dependence from the vector potential, 
but $A$ does still have a ``hidden'' explicit dependence on the arbitrary
reference position. The equation of motion for the reference position is
$F^{\mu\nu}\delta F_{\mu\nu}/\delta X^\lambda_*=0$, which enforces the condition
that the fieldstrength is independent of the reference position.  This 
condition determines the explicit dependence of the potential on $X_*^\mu$.

\section{The string from the dual potential}

The dual potential formalism is useful in many applications since it 
translates the electrodynamics of monopoles into the more familiar 
electrodynamics of charges (and can even allow for the interaction of 
charges and monopoles since electromagnetism is linear).  Here we review
a derivation of the Dirac string from the
dual potential, simplified from versions presented in 
\cite{Deser:1997se,Lechner:1999ga}.  In particular, we start with the
dual fieldstrength and potential and a magnetic current in the absence of 
electric charges (which lead to a Dirac string for the dual fieldstrength)
in order to emphasize that the Poincar\'e duality itself leads to 
the Dirac string for the fieldstrength in the presence of monopoles.

In form notation, the action for the dual electromagnetism with a single
monopole is 
\beq{Sdual}
S = \int \left(-\frac 12 \t F\w\star\t F +\t A\w \star\t\jmath\right) \ ,
\eeq
where $\t A,\t F$ are the dual potential and fieldstrength and $\t\jmath$ is
the monopole current (\ref{current}).  To dualize back to the ``usual'' 
potential in the absence of monopoles, we treat $\t F$ as the independent 
variable and add a Lagrange multiplier term $A d\t F$ to enforce the Bianchi
identity for $\t F$ since $\t A$ does not appear.  Solving the equation of
motion for $\t F$ and substituting gives the usual Maxwell action.

With the monopole current, we must first find a way to eliminate the 
dual potential from (\ref{Sdual}).  We proceed by recalling that any conserved
current can be written as the divergence of a 2-form (in other words,
any co-closed form in Minkowski spacetime is co-exact).
In fact, the Maxwell equation $d\star\t F=\star\t\jmath$ shows that $\t\jmath$
can be written in terms of the monopole's field strength (which also 
satisfies $d\t F=0$).  We have also seen above that the Dirac string
coupling $G$ is another such two form up to the subtraction of a current
along the reference worldline $\M_*$, so we can write
\beqa
S&=&\int \left[-\frac 12 \t F\w\star\t F +\t A\w d\star (gG+\t F_*)\right]
\nonumber\\
&=& \int\left[-\frac 12 \t F\w\star\t F+\t F \w\star(gG+\t F_*)-
A'\w d\t F\right] \!, \;\label{Sdual2}
\eeqa
where $\t F_*$ is the dual fieldstrength of the reference monopole and we
have added the Lagrange multiplier term.  This last term can be rewritten as
$\t F dA'$ up to a total derivative, so the equation of motion is
$\star\t F =\star(gG+\t F_*)-dA'$.  The action is therefore classically
equivalent to
\beqa
S =-\int\frac 12 \left(dA'-\star\t F_*-g\star G\right)
\w \star\left(dA'-\star\t F_*-g\star G\right) \!. \hspace{0.3in}\label{Sdual3}
\eeqa
We recognize $-\star\t F_*=F_*$ as the reference monopole's field strength,
so we define $dA'+F_*=dA$ in terms of a potential $A$ with the appropriate
gauge patches for the reference monopole.  We are left with
$F=dA-g\star G$ and the usual Maxwell action $-\int F\star F/2$ 
including the Dirac string.

\section{An interpretation of the string}\label{s:delta}

We can gain new insight into the Dirac string through the Bianchi identity.
For clarity, we pick a reference frame and work in static gauge for the 
monopole (ie, worldline time $\tau=Y^0$).  In non-relativistic notation,
the Bianchi identity is
\beq{BEbianchi}
\vec\Del\cdot\vec B =g\delta^3(\vec x-\vec X)\ ,\
\vec\Del\times\vec E +\del_t\vec B=-g\del_t\vec X\delta^3(\vec x-\vec X)\, .
\eeq
Our goal is to diagonalize the degrees of freedom, removing the explicit
dependence of $A_\mu$ on the monopole position.  We carry out a formal 
expansion of the monopole current around a static reference monopole at
fixed position $\vec X_*$; therefore, we end up with a static gauge patching
procedure around $\vec X_*$ and will see that the remaining terms can be
organized as a contribution to the fieldstrength.\footnote{This is the 
approach taken in \cite{1609.05904} for D3-branes.}  

We begin by writing 
\beqa
\delta^3(\vec x-\vec X)&=& \delta^3(\vec x-\vec X_*)\label{expanddelta}\\
&+&\sum_{n=1}^\infty
\frac{1}{n!}\delta X^{i_1}\cdots \delta X^{i_n} \Del_{i_1}^*\cdots\Del_{i_n}^*
\delta^3(\vec x-\vec X_*)\ ,\nonumber
\eeqa
where $\vec\Del^*$ is the gradient with respect to $\vec X_*$ and 
$\delta\vec X =\vec X-\vec X_*$.  Converting
one of the derivatives to one with respect to $\vec x$, the terms in the sum
can be written as a divergence, so defining
\beqa
\vec B &=& \vec\Del\times\vec A \label{Bdelta} \\ &-& g\delta\vec X \sum_{n=0}^\infty
\frac{1}{(n+1)!} (\delta\vec X)^{n,i_1\cdots i_{n}} 
(\vec\Del^*)^n_{i_1\cdots i_{n}} \delta^3(\vec x-\vec X_*)\nonumber \eeqa
where $\vec A$ includes the potential of a fixed monopole at $\vec X_*$
solves the scalar equation of (\ref{BEbianchi}).
Subtracting this contribution from the vector part of the Bianchi 
identity similarly extracts a curl from the expansion of the delta function.
We can therefore define
\beqa
\vec E&=& -\vec\Del\Phi-\del_t\vec A +g(\del_t\delta\vec X\times\delta\vec X)
\label{Edelta}\\
&\times& \sum_{n=0}^\infty \frac{n+1}{(n+2)!} (\delta\vec X)^{n,i_1\cdots i_{n}} 
(\vec\Del^*)^n_{i_1\cdots i_{n}} \delta^3(\vec x-\vec X_*)\nonumber
\eeqa
to solve the Bianchi identity.

To confirm that this approach separates the gauge and monopole degrees of 
freedom, we can vary the Maxwell action with respect to $\vec X$ and, with 
some manipulation, find the magnetic Lorentz force equation by treating 
the potentials as independent variables.  As with the Dirac string, the
variation of the action also includes terms proportional to the sourceless
Maxwell equations and derivatives of $\delta^3(\vec x-\vec X_*)$.  

As an alternate approach, we can consider the Dirac string with
$\M_*$ a static worldline at $\vec X_*$.  Then, as we suggested in the section
above, take \N\ at any fixed time to be the line segment from $\vec X_*$ to
$\vec X$. Then we can choose a static gauge\footnote{The argument 
goes through with little change for any function of $\tau$.} 
$t=Y^0(\tau,\sigma)=\tau$ and $\vec Y(\tau,\sigma)=\vec X_*+\sigma\delta
\vec X(t)$ with $0\leq\sigma\leq 1$.  Then the Dirac string coupling becomes
\beqa
G^{0i}&=& \int_0^1d\sigma\, \delta X^i \delta^3(\vec x-\vec X_*-\sigma\delta
\vec X)\ ,\label{Gstatic1}\\
&=&\delta X^i \sum_{n=0}^\infty \frac{1}{n!}(\delta\vec X)^{n,i_1\cdots i_{n}} 
(\vec\Del^*)^n_{i_1\cdots i_{n}} \delta^3(\vec x-\vec X_*)\int_0^1d\sigma\,\sigma^n
\nonumber\eeqa
and similarly
\beqa
G^{ij}&=& (\del_t X^i \delta X^j-\del_t X^j \delta X^i)\sum_{n=0}^\infty 
\frac{1}{n!}(\delta\vec X)^{n,i_1\cdots i_{n}}\nonumber\\
&\times&(\vec\Del^*)^n_{i_1\cdots i_{n}} \delta^3(\vec x-\vec X_*)
\int_0^1d\sigma\, \sigma^{n+1}\label{Gstatic2}
\eeqa
after expanding the delta function.  Carrying out the integral and using
the usual relation between the fieldstrength $F_{\mu\nu}$ and fields 
$\vec E,\vec B$, we find (\ref{Bdelta},\ref{Edelta}).

So the particular solution of the Bianchi identity that we found by 
expanding the delta function around $\vec X_*$ is a particular realization 
of the Dirac string.  As a result, we see that the expansion of the delta 
function separates the explicit dependence of the fieldstrength on the
monopole position from the potential.  This also tells us that the Dirac 
string is a way of treating a monopole's motion as a fluctuation (even a 
large one) around a fixed position. The key difference with Dirac's 
arbitrary string is that this interpretation suggests treating the
embedding coordinates of the string as dependent on the monopole position
along the entire worldsheet.  In fact, we have done this explicitly in
deriving (\ref{Gstatic1},\ref{Gstatic2}).

For future reference, it is useful to give the exact expressions for the
fields with the linear string configuration:
\beqa
\vec E &=& -\vec\Del\Phi-\del_t\vec A +g \left(\del_t\vec X\times \delta
\vec X\right) \int_0^1 d\sigma\,\sigma\delta^3\left(\vec x-\vec X_*-\sigma
\delta\vec X\right)\nonumber\\
\vec B&=& \vec\Del\times\vec A -g\delta\vec X\int_0^1 d\sigma\,\delta^3
\left(\vec x-\vec X_*-\sigma\delta\vec X\right) \ .\label{EBstring}
\eeqa
The displacement $\vec D$ and field $\vec H$ are determined as usual from 
$\vec E,\vec B$ in linear media, including the Dirac string contribution.

\section{The string as a source}

The Dirac string in the fieldstrength automatically solves the Bianchi 
identity, so the Bianchi identity no longer determines the fieldstrength
associated with the dynamic monopole.  Instead, as Dirac realized, the 
string coupling $G$ leads to an effective electric current as a source for $dA$,
the fieldstrength away from the Dirac string.  In particular, 
$d\star dA = -gdG$, so the effective current is $j_{eff}=g\star dG$.

For practical use, we again choose a line segment configuration as in section
\ref{s:delta}.  The effective charge is 
\beqa
\rho_{eff} &=&j_{eff}^0 \label{effcharge}\\
&=& -g\left(\del_t\vec X \times \delta\vec X\right)
\cdot \vec\Del\left[\int_0^1 d\sigma\,\sigma\delta^3(\vec x-\vec X_*-\sigma
\delta\vec X)\right]\ ,\nonumber
\eeqa
where the gradient is with respect to $\vec x$.  Evaluating the effective
current in the static gauge is slightly subtler.  We have
\beqa
j_{eff,i} &=& -g\epsilon_{i0jk}\int d\tau\int_0^1 d\sigma\, (\sigma\del_\tau X^j
\delta X^k)\nonumber\\
&&\times\del_t\left[\delta(t-\tau)\delta^3(\vec x-\vec X_*-\sigma
\delta\vec X)\right]\nonumber\\
&&+g\epsilon_{ij0k}\int_0^1 d\sigma\,\delta X^k\del_j
\delta^3(\vec x-\vec X_*-\sigma\delta\vec X)\ .\eeqa
We must take care to convert the $t$ derivative to a $\tau$ derivative
and integrate by parts where the delta function in time is differentiated.
We are left with
\beqa
\vec\jmath_{eff}&=& g\del_t\left[\left(\del_t\vec X \times \delta\vec X\right)
\int_0^1 d\sigma\,\sigma\delta^3(\vec x-\vec X_*-\sigma
\delta\vec X)\right]\nonumber\\
&&-g\delta\vec X\times \vec\Del\left[\int_0^1 d\sigma\,\delta^3(\vec x-\vec X_*
-\sigma\delta\vec X)\right]\ .
\eeqa

This is worth two comments.  First, the current for arbitrary linear motion
of the monopole, with $\vec X_*$ chosen to lie on the line, is
\beq{effcurrent} \vec\jmath_{eff}= -g\delta\vec X\times
\vec\Del\left[\int_0^1 d\sigma\,\delta^3(\vec x-\vec X_*-\sigma\delta\vec X)
\right]\ .\eeq
Taking $\delta\vec X$ to lie along the $\pm z$ axis, (\ref{effcurrent})
is the surface current in the $\pm\phi$ direction
of an infinitely tightly wound, infinitesimally thin solenoid stretching
from $\vec X_*$ to $\vec X$.  The direction and magnitude of the current are
precisely such that the flux into the solenoid at $\vec X_*$ (which is
spherically symmetric since the solenoid is thin) precisely cancels the
magnetic flux due to the reference monopole at $\vec X_*$.  Similarly, the
flux out of the solenoid at $\vec X$ precisely reproduces the flux of the
dynamical monopole.  Second, the current is conserved, as it must be, if
we allow for singular distributions.  The solenoidal current (\ref{effcurrent})
is clearly conserved, and the other terms take the form
$\rho_{eff}=-\vec\Del\cdot\vec k, \vec\jmath_{eff}=\del_t\vec k$, which is
manifestly conserved.

In media, the effective charge and current follow from $\vec\Del\cdot\vec D$
and $\vec\Del\times\vec H$, so derivatives of the permittivity and
permeability will appear in general.  For piecewise constant dielectric 
constant and permeability, it is natural to define the effective charge and
current in a piecewise manner as well.

Specifying a string configuration that depends explicitly on the monopole
position along the worldsheet gives a well-defined effective current 
from the string.  In principle, we can now solve for the electromagnetic
fields for arbitrary monopole motion as a superposition of the magnetic field
from the fixed reference monopole and the effective current.  The linear
string configuration seems particularly well suited to this type of 
calculation.

\section{Applications}

As we have noted previously, there are technical difficulties in the theory
of electrodynamics with monopoles.  Here we present several possible 
applications in which the linear Dirac string configuration yields 
simplifications.

We indicated above that
one such application is a direct determination of radiation from moving
monopoles, including energy loss in dielectric materials (such as
Cherenkov radiation),\footnote{See \cite{Jackson:1998nia,GeaBancloche:1982fq}
for the energy loss rate in different approximations.} which may be useful
for monopole search experiments.  While an infinite,
arbitrarily moving Dirac string is unwieldy, the linear Dirac string
configuration gives a well-defined current, which is simply a growing or
shrinking solenoid in many contexts. In the presence of materials, as noted,
it is necessary to write the fields nonrelativistically as $\vec E,\vec B$
and include the permittivity and permeability in defining $\vec D,\vec H$.

We have emphasized that the Dirac string formulation separates the gauge
and monopole degrees of freedom; as a result, it provides a basis
for Hamiltonian and therefore quantum treatments.  In this form, 
the introduction of extra unphysical degrees of freedom for the Dirac string 
leads to constraints \cite{Balachandran:1975qc}.  On the other hand,
when the Dirac string takes the linear configuration, the entire string
depends on the monopole position.  In the variation of the action, these
appear through terms proportional to the sourceless Maxwell equations 
(see equation (\ref{eom3})) and are trivial on-shell.  On the other hand,
these terms can contribute off-shell, for example in the path integral.
Understanding how these contribute to the quantum mechanics of monopoles is
an interesting question.  Meanwhile,
the analogous terms for D3-branes (see the discussion below) also play an
important role in the 4-dimensional effective action of type IIB string
theory \cite{1609.05904,andrew}.

The Dirac string formalism straightforwardly extends to curved spacetimes
and higher dimensions,
and the linear configuration for the string becomes a worldsheet with
geodesics as constant-time slices.  A second endpoint to the string on
$\M_*$ therefore allows us to use the Dirac string for monopoles on compact
manifolds.  While the magnetic Gauss law constraint (net magnetic charge
on a compact manifold vanishes) means that any Dirac string can end on 
oppositely charged physical monopoles, an arbitrary reference endpoint 
allows for a cleaner separation of the dynamics of the different monopoles.
Furthermore, it allows us to avoid having multiple Dirac strings end on 
one monopole in the case that the monopoles on a compact manifold carry 
different numbers of magnetic quanta (for example, there are two monopoles
of charge $+1$ and one of charge $-2$).  Again, in the case of 
higher-dimensional monopole-like branes, monopole charge can dissolve into
the flux of other fields, so the Dirac string from a monopole may not 
even have another monopole on which to end, necessitating the reference
endpoint.

Finally, higher-dimensional branes of string theory are magnetic sources
for various rank form fields.  As in electromagnetism, a 
mathematically rigorous treatment considers the potentials as sections,
while a Dirac-like formalism allows the separation of the gauge and brane
degrees of freedom (see \cite{1609.05904}; one of us will detail
this formalism in the forthcoming \cite{andrew}).
The formalism therefore provides an alternative to the democratic
(dual potential) formalism for the derivation of brane equations of motion.
It is particularly helpful for a careful accounting of degrees of
freedom, as needed in dimensional reduction.  Determining the lower-dimensional
effective action is also an off-shell calculation, and extra terms from the
generalized Dirac string worldvolume (which vanish on-shell) are critical 
to account for all the kinetic terms required by supergravity
\cite{1609.05904,andrew}.	

\begin{acknowledgments}
AF would like to thank K.~Dasgupta, N.~Afshordi, and L.~Boyle for interesting
discussions.
BC and AF are supported by the Natural Sciences and
Engineering Research Council of Canada Discovery Grant program.  Part
of this work was supported by the Perimeter Institute
for Theoretical Physics. Research at Perimeter Institute is supported
by the Government of Canada through the Department of Innovation,
Science and Economic Development and by the Province of Ontario
through the Ministry of Research and Innovation.
\end{acknowledgments}

\appendix*
\section{Conventions}
Here we briefly lay out our conventions, including signs.
To start, we take the mostly plus metric convention with $\epsilon_{0123}=+1$.
The Hodge star for a differential $p$-form is given by
$(\star F)_{\mu_1\cdots\mu_{4-p}}=(1/p!)
\epsilon_{\mu_1\cdots\mu_{4-p}}{}^{\nu_1\cdots\nu_p}F_{\nu_1\cdots\nu_p}$, so 
$\star\star F = (-1)^{p(4-p)+1}$.  

With standard conventions (see \cite{Jackson:1998nia,griffiths}), 
the Maxwell equations with magnetic currents included are\beqa
\vec\Del\cdot\vec E =\rho\ &,& \vec\Del\times\vec B -\del_t\vec E=\vec{\jmath}
\nonumber\\
\vec\Del\cdot\vec B =\t\rho\ &,&
\vec\Del\times\vec E +\del_t\vec B=-\vec{\t\jmath}\, ,
\eeqa
where $\rho,\vec\jmath$ are the electric charge and current and
$\t\rho,\vec{\t\jmath}$ are the magnetic.
In relativistic notation, we take $A^\mu=(\Phi,\vec A)$,
$j^\mu=(\rho,\vec\jmath)$ (and likewise for the magnetic current), so
$F_{0i}=-E_i,F_{ij}=\epsilon_{ijk}B_k$.  The Maxwell equations become
\beqa \del_\mu F^{\mu\nu}=-j^\nu , \ \del_{\mu} F_{\nu\lambda}+
\del_\nu F_{\lambda\mu}+\del_\lambda F_{\mu\nu} =-\epsilon_{\mu\nu\lambda\rho}
\t\jmath^\rho , \label{apMax2} \hspace{0.35in}\eeqa
or $d\star F=-\star j$ and $dF = -\star\t\jmath$ in terms of forms. The dual field strength $\t F\equiv \star F$ therefore satisfies the dual
Maxwell equations $d\t F=-\star j$ and $d\star\t F =+\star\t\jmath$.  As a
result, the dual electric current is usually defined as $-\t\jmath$; for
simplicity of comparison, we do \textit{not} introduce this sign.

Finally, we define the Dirac string coupling $G$ as a form integral over the
string worldsheet coordinates $\tau,\sigma$.  We choose the orientation by
taking integration measure $d^2\sigma=d\tau\w d\sigma=-d\sigma\w d\tau$.

\bibliography{diracmonopole}

\end{document}